\documentclass[12pt]{article}
\usepackage{a4}
\usepackage{amsfonts}
\usepackage{amsmath,amssymb}

\def\un#1{\relax\ifmmode\@@underline#1\else
        $\@@underline{\hbox{#1}}$\relax\fi}

\let\du=\du                     

\def\d{\delta}

\def\f{\phi}

\def\h{\eta}

\def\j{\psi}
\def\k{\kappa}

\def\m{\mu}
\def\n{\nu}

\def\q{\theta}

\def\s{\sigma}

\def\L{\Lambda}

\def\ve{\varepsilon}

\def\ce{{\cal E}}

\def\car{{\cal R}}

\def\bo{{\raise-.3ex\hbox{\large$\Box$}}}               
\def\pa{\partial}                                       
\def\TH{{\raise.2ex\hbox{$\displaystyle \bigodot$}\mskip-4.7mu \llap H \;}}
\def\face{{\raise.2ex\hbox{$\displaystyle \bigodot$}\mskip-2.2mu \llap {$\ddot
        \smile$}}}                                      

\def\abs#1{\left| #1\right|}                    
\def\leftrightarrowfill{$\mathsurround=0pt \mathord\leftarrow \mkern-6mu
        \cleaders\hbox{$\mkern-2mu \mathord- \mkern-2mu$}\hfill
        \mkern-6mu \mathord\rightarrow$}
\def\dvec#1{\vbox{\ialign{##\crcr
        \leftrightarrowfill\crcr\noalign{\kern-1pt\nointerlineskip}
        $\hfil\displaystyle{#1}\hfil$\crcr}}}           

\def\frac#1#2{{\textstyle{#1\over\vphantom2\smash{\raise.20ex
        \hbox{$\scriptstyle{#2}$}}}}}                   
\def\sfrac#1#2{{\vphantom1\smash{\lower.5ex\hbox{\small$#1$}}\over
        \vphantom1\smash{\raise.4ex\hbox{\small$#2$}}}} 
\def\bfrac#1#2{{\vphantom1\smash{\lower.5ex\hbox{$#1$}}\over
        \vphantom1\smash{\raise.3ex\hbox{$#2$}}}}       
\def\afrac#1#2{{\vphantom1\smash{\lower.5ex\hbox{$#1$}}\over#2}}    

\def\[{\lfloor{\hskip 0.35pt}\!\!\!\lceil}
\def\]{\rfloor{\hskip 0.35pt}\!\!\!\rceil}

\def\du#1#2{_{#1}{}^{#2}}

\def\un{\underline}
\def\fracmm#1#2{{{#1}\over{#2}}}

\def\low#1{{\raise -3pt\hbox{${\hskip 0.75pt}\!_{#1}$}}}

\newskip\humongous \humongous=0pt plus 1000pt minus 1000pt
\def\caja{\mathsurround=0pt}
\def\eqalign#1{\,\vcenter{\openup2\jot \caja
        \ialign{\strut \hfil$\displaystyle{##}$&$
        \displaystyle{{}##}$\hfil\crcr#1\crcr}}\,}
\newif\ifdtup

\newcommand{\be}{\begin{equation}}
\newcommand{\ee}{\end{equation}}
\newcommand{\nbe}{\begin{equation*}}
\newcommand{\nee}{\end{equation*}}

\newcommand{\lb}{\label}

\begin{document}

\thispagestyle{empty}

{\hbox to\hsize{
\vbox{\noindent February 2011 \hfill version 2}}}

\noindent
\vskip2.0cm
\begin{center}

{\large\bf     COSMOLOGICAL PROPERTIES OF \\
\vglue.1in   A GENERIC $\car^2$-SUPERGRAVITY~\footnote{Supported
 in part by the Japanese Society for Promotion of Science (JSPS)}}
\vglue.3in

          Sergei V. Ketov~${}^{a,b}$ and Natsuki Watanabe~${}^a$
\vglue.1in

${}^a$ {\it Department of Physics, Graduate School of Science, 
   Tokyo Metropolitan University, Hachioji-shi, Tokyo 192-0397, Japan}\\
${}^b$ {\it Institute for Physics and Mathematics of Universe, The 
           University of Tokyo, Kashiwa-shi, Chiba 277-8568, Japan}
\vglue.1in
ketov@phys.se.tmu.ac.jp, watanabe-natsuki1@ed.tmu.ac.jp
\end{center}

\vglue.3in

\begin{center}
{\Large\bf Abstract}
\end{center}
\vglue.1in

\noindent We investigate in detail the structure of the simplest non-trivial
$F(\car)$-supergravity model, whose $F$-function is given by a generic {\it 
quadratic} polynomial in terms of the scalar supercurvature $\car$. This 
toy-model admits a fully explicit derivation of the corresponding 
$f(R)$-gravity functions. We apply the stability requirements for selecting 
the physical $f(R)$-gravity functions, and discuss the phenomenological 
prospects of $F(\car)$-supergravity in its application to cosmology. 

\newpage

\section{Introduction}

Unknown inflaton and unification of cosmological inflation with High-Energy 
Physics remain the outstanding problems beyond the Standard Models of 
elementary particles and cosmology. One of the easy ways of theoretical 
realization of an inflationary universe is given by popular theories of $f(R)$ 
gravity whose Lagrangian is a function $f(R)$ of the scalar curvature  $R$ in 
four space-time dimensions  (see refs.~{\cite{fr,stu} for a recent review). The
 use of those theories in the inflationary cosmology was pioneered by 
Starobinsky \cite{star,star3}.

An $f(R)$ theory of gravity is classically equivalent to the certain 
scalar-tensor theory of gravity \cite{eq,bac,ma,dh}. In $f(R)$-gravity 
a dynamics of the spin-2 part of metric is not modified, but there is the extra
 propagating scalar field (called scalaron) given by the conformal mode of the 
metric (over Minkowski or (anti)de Sitter vacuum). That scalar field plays the 
role of inflaton in the inflationary models based on $f(R)$ gravity. The $f(R)$
 gravity emerged as the phenomenological approach unrelated to any fundamental 
theory of gravity and without a connection to High-Energy Physics. We believe 
that such connection may be established via embedding and extending 
$f(R)$-gravity to supergravity, as the first step. Supergravity is more
fundamental than gravity because supergravity is the theory of {\it local}
supersymmetry, while local supersymmetry implies general covariance. 

In our recent papers \cite{our1,our2,our3,our4,ks} we proposed the new 
supergravity theory that we call $F(\car)$-supergravity. It can be considered 
as the $N=1$ locally supersymmetric extension of $f(R)$ 
gravity.~\footnote{Another (unimodular) $F(R)$ supergravity theory was proposed
 in ref.~\cite{nish}.} Supergravity is well-motivated in High-Energy Physics 
theory beyond the Standard Model of elementary particles. Supergravity is also 
the low-energy effective action of Superstrings that is the theory of
Quantum Gravity.~\footnote{Some applications of $F(\car)$ supergravity to Loop
Quantum Gravity were given in ref.~\cite{gky}.} Unlike the $f(R)$ theories
 of gravity, the $F(\car)$ supergravity is more constrained by local 
supersymmetry and consistency. Moreover, the manifestly supersymmetric 
construction of $F(\car)$ supergravity in superspace 
\cite{our1,our2,our3} leads to a {\it chiral\/} action in curved $N=1$ 
superspace, which may be naturally stable against quantum corrections that are 
usually given by full superspace integrals. Our supersymmetric extension of 
$f(R)$ gravity is non-trivial because the $F(\car)$ supergravity auxiliary 
fields do not propagate (this feature is called the auxiliary freedom 
\cite{gat}). Similarly to $f(R)$-gravity, the (complex) superconformal mode of 
the supergravity supervielbein (over Minkowski or anti-de Sitter vacuum) 
becomes dynamical in $F(\car)$ supergravity. As was demonstrated in 
ref.~\cite{our1}, an $F(\car)$ supergravity is classically equivalent to the 
standard $N=1$ Poincar\'e supergravity coupled to the dynamical chiral 
superfield, whose (non-trivial) K\"ahler potential and superpotential are 
dictated by a chiral (holomorphic) function. The extra dynamical chiral 
superfield is just the superconformal mode of the supervielbein, or a complex 
scalaron. As was argued in ref.~\cite{our1}, the leading complex scalar field 
component of the chiral superfield (the superscalaron) may be identified with 
the dilaton-axion field in (non-perturbative) Superstrings/M-Theory.
 
The component structure of $F(\car)$ supergravity is very complicated, and 
some of its general features were studied in refs.~\cite{our3,our4}. In
particular, the first explicit derivation of a real bosonic function $f(R)$ out
 of the supergravity (holomorphic) function $F(\car)$ was given in 
ref.~\cite{our4}. In ref.~\cite{ks} the natural embedding of the 
$(R+R^2)$-inflationary model into $F(\car)$ supergravity was found, which gives
 a simple and viable realization of chaotic inflation \cite{chaot} in 
supergravity. In this paper we further extend the results of ref.~\cite{our4} 
to the case of $F$-function given by a {\it generic quadratic} polynomial in 
terms of the scalar supercurvature, and analyze all possible solutions to the 
corresponding bosonic $f(R)$-gravity functions. We also apply the stability 
requirements to select those of them which are physical.

Our paper is organized as follows. In sec.~2 we briefly recall the superspace 
construction of $F(\car)$ supergravity, and provide the algebraic equations for
 the auxiliary fields. In sec.~3 we define our model of $F(\car)$ supergravity,
 and explicitly derive the corresponding bosonic functions $f(R)$. The 
stability conditions are applied in sect.~4. Possible physical applications are
 discussed in sec.~5. Our conclusion and outlook is sec.~6.

\section{$F(\car)$ supergravity and its auxiliary scalars}

A concise and manifestly supersymmetric description of supergravity is given
by superspace. We refer the interested reader to the textbooks 
\cite{ss1,ss2,ss3} for details. Here we use the units $c=\hbar=1$ and 
$\k=M_{\rm Pl}^{-1}$ in terms of the (reduced) Planck mass $M_{\rm Pl}$, with 
the spacetime signature $(+,-,-,-)$. Our basic notation of General Relativity 
coincides with that of ref.~\cite{landau}.

The most succinct formulation of $F(\car)$ supergravity exist in a chiral
4D, $N=1$ superspace where  it is defined by the action \cite{our1}
\be  \lb{act}
 S = \int d^4x d^2\q\, \ce F(\car) + {\rm H.c.}
\ee
in terms of a holomorphic function $F(\car)$ of the covariantly-chiral scalar
curvature superfield $\car$, and the chiral superspace density $\ce$. The 
chiral
$N=1$ superfield $\car$ has the scalar curvature $R$ as the field coefficient
at its $\q^2$-term. The chiral superspace density $\ce$ (in a WZ gauge) reads
\be \lb{cde}
\ce = e \left( 1- 2i\q\s_a\bar{\j}^a +\q^2 B\right) 
\ee
where $e=\sqrt{-g}$, $\j^a$ is gravitino, and $B=S-iP$ is the complex scalar 
auxiliary field (it does not propagate in the theory (\ref{act}) despite of the
apparent presence of the higher derivatives). The theory (\ref{act}) is 
classically 
equivalent to the standard $N=1$ Poincar\'e supergravity minimally coupled to 
the chiral 
scalar superfield, via the supersymmetric Legendre-Weyl-K\"ahler transform 
\cite{our1,our2}. The chiral scalar superfield, given by the superconformal 
mode of the
supervielbein, becomes dynamical in a generic $F(\car)$ supergravity.   

As regards a {\it large-scale} evolution of the FRLW Universe in terms of its 
scale factor, it is the {\it scalar} curvature dependence of any gravitational 
effective action that plays the most relevant role there. Similarly, as regards
 any supergravitational effective action, the evolution of the FRLW scale 
factor is largely determined by a dependence of the gravitational superfield 
effective action upon the scalar supercurvature $\car$.

A bosonic $f(R)$ gravity action is given by \cite{fr,stu}
\be \lb{mgrav}
 S_{\rm f} = \int d^4x \,\sqrt{-g}\, f(R) \ee
in terms of the real function $f(R)$ of the scalar curvature $R$. The relation 
between the master chiral superfield  function $F(\car)$ in eq.~(\ref{act}) and
 the corresponding bosonic function $f(R)$ in eq.~(\ref{mgrav}) can be 
established by appplying the standard formulae of superspace \cite{ss1,ss2,ss3}
 and ignoring the fermionic contributions.  As a result \cite{our2,our3,our4}, 
one gets a bosonic Lagrangian in the form
\be \lb{expa}
(-g)^{-1/2}L_{\rm bos}\equiv f(R,\tilde{R};X,\bar{X})=
F'(\bar{X}) \left[ \frac{1}{3}R_* +4\bar{X}X \right] +3X F(\bar{X})+{\rm 
H.c.}  \ee
where the primes denote differentiation, and we have also introduced the 
notation
\be \lb{not1} X=\fracmm{\k}{3}B \qquad {\rm and} \qquad 
R_*=R-\frac{i}{2}\ve^{abcd}R_{abcd}\equiv R + i\tilde{R}~.\ee
The $\tilde{R}$ does not vanish in $F(\car)$ supergravity, and it represents 
an axion, the pseudo-scalar superpartner of real scalaron (inflaton) 
in our construction \cite{our1,our2,our3,our4,ks}.  

Varying eq.~(\ref{expa}) with respect to the complex auxiliary fields $X$ and 
$\bar{X}$,
\be \lb{auxe}
 \fracmm{\pa L_{\rm bos}}{\pa X} =  
\fracmm{\pa L_{\rm bos}}{\pa \bar{X}} = 0~~,
\ee
 gives rise to the algebraic equations on the auxiliary fields,
\be\lb{aux1}
3\bar{F}+X(4\bar{F}'+7F')+4\bar{X}XF'' +\frac{1}{3}F''R_*=0
\ee
and its conjugate
\be \lb{aux2}
3F+\bar{X}(4F'+7\bar{F}')+4\bar{X}X\bar{F}'' +\frac{1}{3}\bar{F}''\bar{R}_*=0
\ee
where $F=F(X)$ and $\bar{F}=\bar{F}(\bar{X})$. The algebraic equations 
(\ref{aux1}) and (\ref{aux2}) cannot be explicitly solved for $X$ in a generic 
$F(\car)$ supergravity.

The stability conditions in $f(R)$-gravity are well known
\cite{fr,stu}, and in our notation they read
\be \lb{csta} 
 f'(R) < 0 
\ee
and
\be \lb{qsta} 
 f''(R) > 0
\ee
The first (classical stability) condition (\ref{csta}) is related to the sign 
factor in front of the Einstein-Hilbert term (linear in $R$) in the 
$f(R)$-gravity action, and it ensures that graviton is not a ghost. The second 
(quantum stability) condition (\ref{qsta}) ensures that scalaron is not a 
tachyon.

Being interested in the bosonic $f(R)$-gravity action that follows from 
eq.~(\ref{act}), we set both gravitino and axion to zero, which also implies 
$R_*=R$ and a {\it real} $X$. 

In $F(R)$ supergravity, eq.~(\ref{csta}) is to be replaced by a stronger 
condition \cite{ks},
\be \lb{cs}
F'(X) < 0 
\ee
It is easy to verify that eq.~(\ref{csta}) follows from eq.~(\ref{cs}) because 
of eq.~(\ref{auxe}). Equation (\ref{cs}) also guarantees the classical 
stability of the bosonic $f(R)$ gravity embedding into the full $F(\car)$ 
supergravity against small fluctuations of the axion field \cite{ks}.

\section{Our toy-model}

First, we recall that the standard (pure) supergravity \cite{ss1,ss2,ss3} 
is reproduced in our approach by simply taking 
\be \lb{case0} F''=0 \qquad {\rm or,~equivalently,} \qquad 
F(\car)=f_0-\fracmm{1}{2}f_1\car~,\ee
with some (complex) constants $f_0$ and $f_1$, where ${\rm Re}f_1>0$.  
Then eqs.~(\ref{aux1}) and (\ref{aux2}) are easily solved by
\be \lb{sol0}
X =\fracmm{3f_0}{5({\rm Re}f_1)} \ee
Substituting this solution back into the Lagrangian (\ref{expa}) yields
\be \lb{sug}
(-g)^{-1/2}L_{\rm bos} = -\frac{1}{3}({\rm Re}f_1)R 
+\fracmm{9\abs{f_0}^2}{5({\rm Re}f_1)}
\equiv -\fracmm{1}{2}M^2_{\rm Pl}R -\L \ee
where we have identified
\be \lb{sugf}
  {\rm Re}f_1= \fracmm{3}{2}M^2_{\rm Pl}\qquad {\rm and}\qquad
 \L = \fracmm{-9\abs{f_0}^2}{5({\rm Re}f_1)}=
 \fracmm{-6\abs{f_0}^2}{5M^2_{\rm Pl}}
 \ee
As is clear from the above equations, the cosmological constant in the
{\it standard} pure supergravity is always {\it zero or negative}, as is 
required by local supersymmetry. Since we are not interested in the standard 
supergravity, we assume that $F''\neq 0$ in what follows.

Let's now investigate the simplest non-trivial Ansatz ($F''=const.\neq 0$)
for the $F(R)$ supergravity function in the form
\be \lb{an}
 F(\car) = f_0   -\fracmm{1}{2}f_1 \car + \fracmm{1}{2}f_2 \car^2 
\ee
with three coupling constants $f_0$, $f_1$ and $f_2$. We will take all of 
them to be real, since we will ignore this potential source of $CP$-violation
in what follows.  As regards the mass dimensions of the quantities introduced, 
we have
\be \lb{dims}
[F]=[f_0]=3~, \quad [R]=[f_1]=2~, \quad{\rm and}\quad [\car]=[f_2]=1 
\ee

The bosonic Lagrangian (\ref{expa}) with the function (\ref{an}) reads
\be \lb{simple}
(-g)^{-1/2}L_{\rm bos}= 11 f_2 X^3 - 7f_1X^2 
+ \left( \frac{2}{3}f_2R +6f_0\right)X -\frac{1}{3}f_1R
\ee
Hence, the auxiliary field equation (\ref{auxe}) takes the form of
 a {\it quadratic} equation,
\be \lb{xquad}
\frac{33}{2}f_2 X^2 - 7f_1X + \frac{1}{3}Rf_2 + 3f_0 =0
\ee
whose solution is given by
\be \lb{xquads}
 X_{\pm} =  \fracmm{7}{3\cdot 11} \left[ \fracmm{f_1}{f_2} \pm 
\sqrt{ \fracmm{2\cdot 11}{7^2} (R_{\rm max}-R)}\right]
\ee
where we have introduced the {\it maximal\,} scalar curvature ({\sl cf.} 
refs.~\cite{bi,gk})
\be \lb{max}
R_{\rm max} = \fracmm{7^2}{2\cdot 11}\fracmm{f^2_1}{f^2_2} 
-3^2\fracmm{f_0}{f_2}
\ee
Equation (\ref{xquads}) obviously implies the automatic bound \cite{our4} 
\be \lb{bound} 
R<R_{\rm max}
\ee
It is worth mentioning that eq.~(\ref{bound}) is better interpreted as the 
{\it lower} (or AdS) bound on the scalar curvature $(-R)$. For example, in our
notation, a de-Sitter space has a negative (constant) scalar curvature 
$R_{\rm dS}<0$, whereas an anti-de-Sitter space has a positive (constant)
scalar curvature, $R_{\rm AdS}>0$. It is therefore natural to demand 
$R_{\rm max}>0$ (or, equivalently, $-R_{\rm max}<0$), in order to allow a flat
space $(R=0)$ too. It yields
\be \lb{neq}
198 f_0f_2 < (7f_1)^2
\ee

Substituting the solution (\ref{xquads}) back into eq.~(\ref{simple}) yields 
the corresponding $f(R)$-gravity Lagrangian (after a tedious but 
straightforward calculation) with
\be \lb{flag}
\eqalign{
f_{\pm}(R) = & \fracmm{2\cdot 7}{11}\fracmm{f_0f_1}{f_2} 
-\fracmm{2\cdot 7^3}{3^3\cdot 11^2}
\fracmm{f^3_1}{f^2_2} \cr
&  -\fracmm{19}{3^2\cdot 11} f_1R \mp \sqrt{ \fracmm{2}{11}}\left(
\fracmm{2^2}{3^3}f_2\right)
 \left( R_{\rm max}-R \right)^{3/2}  \, } 
\ee
Expanding eq.~(\ref{flag}) into power series of $R$ yields
\be \lb{taylor}
f_{\pm}(R) = -\L_{\pm} - a_{\pm}R +b_{\pm}R^2 +{\cal O}(R^3)
\ee
whose coefficients are given by 
\be \lb{cosc}
\L_{\pm} = \fracmm{2\cdot 7}{3^2\cdot 11}f_1 \left(
R_{\rm max} - \fracmm{7^2}{2\cdot 3\cdot 11} \fracmm{f_1^2}{f^2_2}\right)
\pm \sqrt{\fracmm{2}{11}}\left(\fracmm{2^2}{3^3}f_2\right)R^{3/2}_{\rm max}
\ee
\be \lb{einhil}
a_{\pm} = \fracmm{19}{3^2\cdot 11}f_1 \mp \sqrt{\fracmm{2}{11}R_{\rm max}}
\left(\fracmm{2}{3^2}f_2\right)
\ee
and
\be \lb{2ndc}
b_{\pm} = \mp \sqrt{\fracmm{2}{11R_{\rm max}}} \left( \fracmm{f_2}{2\cdot 3^2}
\right) 
\ee

Those equations greatly simplify when $f_0=0$. One finds \cite{our4}
\be \lb{fgro}
f^{(0)}_{\pm}(R) = \fracmm{-5\cdot 17 M^2_{\rm Pl} }{2\cdot 3^2\cdot 11} R
+ \fracmm{2\cdot 7}{3^2\cdot 11}M^2_{\rm Pl} 
\left(R - R_{\rm max} \right)\left[ 1\pm \sqrt{1-R/R_{\rm max} } \; \right] 
\ee
where we have chosen
\be \lb{sol1}
f_1= \fracmm{3}{2}M^2_{\rm Pl}
\ee
in order to get the standard normalization of the Einstein-Hilbert term that 
is linear in $R$. Then, in the limit $R_{\rm max}\to +\infty$, both functions 
$f^{(0)}_{\pm}(R)$ reproduce General Relativity. In another limit $R\to 0$, one
finds a {\it vanishing} or {\it positive} cosmological constant \cite{our4},
\be \lb{cc}
\L^{(0)}_- =0 \qquad {\rm and}\qquad  
\L^{(0)}_+ = \fracmm{2^2\cdot 7}{3^2\cdot 11}M^2_{\rm Pl}R_{\rm max}
\ee

\section{Stability conditions}

The stability conditions are given by eqs.~(\ref{csta}), (\ref{qsta}) and 
(\ref{cs}), while the 3rd condition implies the 2nd one (sec.~2). In our case 
(\ref{flag}) we have
\be \lb{1der}
f'_{\pm}(R) = -\fracmm{19}{3^2\cdot 11} f_1 \pm \sqrt{\fracmm{2}{11}}\left(
\fracmm{2}{3^2}f_2 \right)\sqrt{R_{\rm max}-R}~ < 0
\ee
and
\be \lb{2der}
f''_{\pm}(R) = \mp \left( \fracmm{f_2}{3^2}\right)\sqrt{
\fracmm{2}{11(R_{\rm max}-R)}}~>0
\ee
while eqs.~(\ref{cs}), (\ref{an}) and  (\ref{xquads}) yield 
\be \lb{csf}
\pm \sqrt{ \fracmm{2\cdot 11}{7^2}(R_{\rm max}-R)}~ < \fracmm{19}{2\cdot 7}
\fracmm{f_1}{f_2}
\ee

It follows from eq.~(\ref{2der}) that 
\be
f_2^{(+)} < 0 \qquad {\rm and}\qquad f_2^{(-)} > 0 
\ee
Then the stability condition (\ref{qsta}) is obeyed for any value of $R$.

{\it As regards the $(-)$-case,} there are {\it two} possibilities depending 
upon the
sign of $f_1$. Should $f_1$ be {\it positive}, all the remaining stability 
conditions 
are automatically satisfied, ie. in the case of both $f_2^{(-)}>0$ and 
$f_1^{(-)}>0$.

Should $f_1$ be {\it negative,} $f_1^{(-)}<0$, we find that the remaining 
stability
conditions (\ref{1der}) and (\ref{csf}) are {\it the same}, as they should, 
while they are both given by
\be \lb{ssc}
R < R_{\rm max} -\fracmm{19^2}{2^3\cdot 11}\fracmm{f_1^2}{f_2^2} =
-\fracmm{3\cdot 5}{2^3\cdot 11}\fracmm{f_1^2}{f_2^2} -3^2\fracmm{f_0}{f_2}
\equiv R_{\rm max}^{\rm ins}
\ee

{\it As regards the $(+)$-case,} eq.~(\ref{csf}) implies that $f_1$ should be 
{\it negative}, $f_1<0$, whereas then eqs.~(\ref{1der}) and (\ref{csf}) result 
in {\it the same} condition (\ref{ssc}) again.

Since $R_{\rm max}^{\rm ins} < R_{\rm max}$, our results imply that the 
instability happens {\it before} $R$ reaches $R_{\rm max}$ in all cases with 
negative $f_1$.

As regards the particularly simple case (\ref{fgro}), the stability conditions 
allow us to choose the lower sign only.~\footnote{The same sign was also chosen
 in  ref.~\cite{our4}, though without giving an explanation.} 

A different example arises with a negative $f_1$. When choosing the lower sign 
(ie. a positive $f_2$) for definiteness, we find
\be \lb{mpc}
\eqalign{
f_{-}(R) = & -\fracmm{2\cdot 7}{11}f_0 \abs{\fracmm{f_1}{f_2}} +
\fracmm{2\cdot 7^3}{3^3\cdot 11^2}\abs{\fracmm{f^3_1}{f^2_2}} \cr
&  +\fracmm{19}{3^2\cdot 11} \abs{f_1}R + 
\sqrt{ \fracmm{2}{11}}\left(\fracmm{2^2}{3^3}f_2\right)
 \left( R_{\rm max}-R \right)^{3/2}  \, } 
\ee
Demanding the standard normalization of the Einstein-Hilbert term in this case 
implies
\be \lb{norm}
R_{\rm max} = \fracmm{3^4\cdot 11}{2^3 f^2_2}
\left( \fracmm{M^2_{\rm Pl}}{2}+\fracmm{19}{3^2\cdot 11}\abs{f_1}\right)^2
\ee
where we have used eq.~(\ref{einhil}). It is easy to verify by using 
eq.~(\ref{cosc}) 
that the cosmological constant is always {\it negative} in this case, and the 
instability bound (\ref{ssc}) is given by
\be \lb{insb}
R_{\rm max}^{\rm ins} = \fracmm{3^4\cdot 11 M^2_{\rm Pl}}{2^3f_2^2}\left(
\fracmm{M^2_{\rm Pl}}{2^2} + \fracmm{19\abs{f_1}}{3^2\cdot 11}\right) 
< R_{\rm max}
\ee

\section{Some applications}

The $f_{-}(R)$ function of eq.~(\ref{flag}) can be rewritten to the form
\be \lb{cho}
f(R)= \fracmm{7^3}{3^3\cdot 11^2}\fracmm{f_1^3}{f_2^2}
-\fracmm{2\cdot 7}{3^2\cdot 11}
f_1R_{\rm max}-\fracmm{19}{3^2\cdot 11}f_1R
 + f_2\sqrt{ \fracmm{2^5}{3^6\cdot 11}} ( R_{\rm max}-R)^{3/2}
\ee
where we have used eq.~(\ref{max}). There are {\it three} physically different 
regimes:

(i) the {\it high-curvature regime,} 
$R<0$ and $\abs{R}\gg R_{\rm max}$. Then eq.~(\ref{cho}) implies
\be 
f(R) \approx -\L_h - a_hR +c_h\abs{R}^{3/2}
\ee
whose coefficients are given by
\be
\eqalign{
\L_h = &~ \fracmm{2\cdot 7}{3^2\cdot 11}f_1R_{\rm max} - 
\fracmm{7^3}{3^3\cdot 11^2}\fracmm{f_1^3}{f_2^2}~~, \cr
a_h = &~ \fracmm{19}{3^2\cdot 11}f_1~~, \cr
c_h = &~ \sqrt{ \fracmm{2}{ 11 } } \left( \fracmm{2^2}{3^3}f_2 \right) \cr}
\ee

(ii) the {\it low-curvature regime,} $\abs{R/R_{\rm max}}\ll 1$. Then 
eq.~(\ref{cho}) implies
\be 
f(R) \approx -\L_l - a_lR~,
\ee
whose coefficients are given by
\be \lb{lcc}
\eqalign{
\L_l = &~ \L_h -\sqrt{ \fracmm{2R^3_{\rm max}}{11} }\left( 
\fracmm{2^2}{3^3}f_2\right)~~, 
\cr
a_l = &~ a_h +\sqrt{\fracmm{2R_{\rm max}}{11}}\left(\fracmm{2}{3^2}f_2\right)=
a_{-}=\fracmm{M^2_{\rm Pl}}{2}~~, \cr }
\ee
where we have used eq.~(\ref{einhil}).

(iii) the {\it near-the-bound regime} (assuming that no instability happens 
before it), $R=R_{\rm max}+\d R$, $\d R <0$, and $\abs{\d R/R_{\rm max}}\ll 1$.
 Then eq.~(\ref{cho})  implies 
\be f(R) \approx - \L_b +a_b\abs{\d R} +c_b \abs{\d R}^{3/2}
\ee
whose coefficients are
\be 
\eqalign{
\L_b = &~ \fracmm{1}{3}f_1R_{\rm max} - 
\fracmm{7^3}{3^3\cdot 11^2}\fracmm{f_1^3}{f_2^2}~~, \cr
a_b = &~ a_h~~, \cr
c_b = &~ \sqrt{ \fracmm{2}{11} }\left( \fracmm{2^2}{3^3}f_2\right)
\cr }
\ee

The cosmological dynamics may be either directly derived from the gravitational
equations of motion in the $f(R)$-gravity with a given function $f(R)$, or just
read off from the form of the corresponding scalar potential of a scalaron 
(see below). For example, as was demonstrated in ref.~\cite{our4} for the 
special case $f_0=0$,~\footnote{See also ref.~\cite{ks} for another
example.} a cosmological expansion is possible in the regime (i) towards the
regime (ii), and then, perhaps, to the regime (iii) unless an instability 
occurs.

However, one should be careful since our toy-model (\ref{an}) does not pretend 
to be viable in the low-curvature regime, eg., for the present Universe. 
Nevertheless, if one wants to give it some physical meaning there, by 
identifying it with General Relativity, then 
one should also fine-tune the cosmological constant $\L_l$ in eq.~(\ref{lcc}) 
 to be ``small'' and positive. We find that it amounts to 
\be \lb{grb}
R_{\rm max} \approx \fracmm{3^4\cdot 7^2\cdot 11}{2^5\cdot 19^2}
\fracmm{M^4_{\rm Pl}}{f_2^2}
\equiv R\low{\L=0}
\ee 
with the actual value of $R_{\rm max}$ to be ``slightly'' above of that bound, 
$R_{\rm max}>R\low{\L=0}$. It is also posssible to have the vanishing 
cosmological constant, $\L_l=0$, when choosing $R_{\rm max}=R\low{\L=0}$. It is
 worth mentioning that it relates the values of $R_{\rm max}$ and $f_2$.

As is well known \cite{eq,bac,ma,llbook,kkw1}, an $f(R)$ gravity theory 
(\ref{mgrav}) is  classically equivalent to a scalar-tensor gravity having the 
action 
\be \lb{st}
S[g_{\m\n},\f] =  \int d^4x\, \sqrt{-g}\left\{ \fracmm{-R}{2\k^2}
+\fracmm{1}{2}g^{\m\n}\pa_{\m}\f\pa_{\n}\f - V(\f) \right\} \ee
in terms of the scalaron field $\f(x)$ with the scalar potential $V(\f)$.  The 
equivalence is established via a Legendre-Weyl transform \cite{eq,ma,kkw1}. 
In our notation we have~\footnote{Compared to ref.~\cite{kkw1}, we changed 
here $y\to -y$.}
\be \lb{nota} 
f(R)=Z(e^y)-Re^y~,\quad R=Z'(e^y)~,\quad -f'(R)=e^y~,\qquad 
y= \sqrt{\fracmm{2}{3}} \fracmm{\f}{M_{\rm Pl}}
\ee
so that the scalar potential is given by \cite{kkw1}
\be \lb{spo}
 V(y) = -\fracmm{1}{2}M^2_{\rm Pl}e^{-2y}Z(e^y)
\ee

It is worth noticing that the stability condition (\ref{csta}) implies an 
invertibility of the 3rd equation (\ref{nota}), ie. $R=R(y)$.

For instance, the simplest Starobinsky model of chaotic inflation 
\cite{star,ks} corresponds to 
\be \lb{eff}
 f_S(R) = -\fracmm{1}{2}M^2_{\rm Pl} \left( R-\fracmm{R^2}{6M^2}\right)
\ee
where the mass parameter $M$ coincides with the inflaton mass. The 
corresponding inflaton scalar potential (\ref{spo}) is given by 
\cite{llbook,kkw1}
\be \lb{starsp}  V(y) = V_0 \left( e^{-y}-1\right)^2 \ee
where $V_0=\fracmm{3}{4}M^2_{\rm Pl}M^2$. 
The constant term in eq.~(\ref{starsp}) is the vacuum energy that drives 
inflaton towards the minimum of its scalar potential (so that the inflation has
 an end). In terms of the equivalent scalar-tensor gravity (\ref{st}) with the 
scalar ponential (\ref{starsp}) the standard slow-roll parameters \cite{llbook}
 are given by \cite{kkw1}
\be \lb{apeps}
\ve = \fracmm{1}{2} M^2_{\rm Pl} \left( \fracmm{V'}{V}\right)^2
= \fracmm{4e^{-2y}}{3\left( e^{-y}-1 \right)^2} =
\fracmm{3}{4N^2_e} +{\cal O}\left( \fracmm{\ln^2 N_e}{N^3_e}\right) 
\ee
and
\be \lb{apeta}
\eta =  M^2_{\rm Pl} \fracmm{V''}{V}  = 
\fracmm{4e^{-y}(2e^{-y}-1)}{3\left(e^{-y}-1 \right)^2} =
 - \fracmm{1}{N_e} + \fracmm{3\ln N_e}{4N_e^2} +\fracmm{5}{4N^2_e}
+{\cal O}\left( \fracmm{\ln^2 N_e}{N^3_e}\right) 
\ee
where the primes denote the derivatives with respect to the inflaton field 
$\f$, and the e-foldings number $N_e$ reads 
\be \lb{efol}
N_e = \int^{t_{\rm end}}_t H dt \approx 
\fracmm{1}{M^2_{\rm Pl}} \int^{\f}_{\f_{\rm end}} \fracmm{V}{V'} d\f
\approx \fracmm{3}{4}\left( e^y-y\right)-1.04
\ee
The theoretical values of the CMB spectral indices \cite{llbook},  
\be \lb{sind}
n_s  = 1+2\h -6\ve \quad {\rm and} \quad r=16\ve ~~,
\ee
in the case of the Starobinsky model are given by \cite{kkw1}
\be \lb{apns}
 n_s =1 - \fracmm{2}{N_e} + \fracmm{3\ln N_e}{2N_e^2} -\fracmm{2}{N^2_e}
+{\cal O}\left( \fracmm{\ln^2 N_e}{N^3_e}\right) 
\ee
and
\be \lb{rho}
r = \fracmm{12}{N^2_e} +{\cal O}\left( \fracmm{\ln^2 N_e}{N^3_e}\right) 
\ee
They agree with the old estimates \cite{mchi}, as well as the most
recent WMAP7 data \cite{wmap7}, when choosing $N_e\approx 54$ and
\be \lb{scale} \fracmm{M}{M_{\rm Pl}}= (3.5\pm 1.2)\cdot 10^{-5}
\ee
so that the Starobinsky inflationary scenario \cite{star} is still viable.

The particular $\car^2$-supergravity model (with $f_0=0$) was introduced in 
ref.~\cite{our4} in an attempt to get viable embedding of the Starobinsky model
 into $F(\car)$-supergravity. However, it failed because, as was found in 
ref.~\cite{our4}, the higher-order curvature terms cannot be ignored in 
eq.~(\ref{fgro}), ie. the $R^n$-terms with $n\geq 3$ are not small enough 
against the $R^2$-term.~\footnote{The possibility of destabilizing the 
Starobinsky inflationary scenario by the terms with higher powers of the scalar
 curvature, in the context of $f(R)$-gravity, was noticed earlier in 
refs.~\cite{ma2,bma}.} The most general Ansatz (\ref{an}), which is merely
{\it quadratic} in the supercurvature, does not help for that purpose either.

For example, the full $f(R)$-gravity function $f_-(R)$ in eq.~(\ref{fgro}), 
which we derived from our $\car^2$-supergravity, gives rise to the inflaton 
scalar potential
\be \lb{expot}
V(y) = V_0 \left( 11e^y +3\right)\left(e^{-y}-1\right)^2
\ee
where $V_0= (3^3/2^6)M^4_{\rm Pl}/f_2^2$. The corresponding inflationary  
parameters 
\be \lb{exeps}
\ve (y) = \fracmm{1}{3} \left[ \fracmm{ e^y\left( 11+ 11e^{-y} +6 e^{-2y}
\right) }{ (11e^y+3)
(e^{-y}-1)} \right]^2  \geq \fracmm{1}{3} \ee
and
\be \lb{exeta}
\eta(y)= \fracmm{2}{3} \fracmm{\left(11 e^y+5e^{-y}+12 e^{-2y}\right) }{ 
(11e^y +3) (e^{-y}-1)^2 } \geq \fracmm{2}{3}
\ee
are not small enough for matching the WMAP observational data. A solution to 
this problem was found in ref.~\cite{ks} by adding to eq.~(\ref{an}) an extra 
term that is {\it cubic} in the supercurvature, with a large dimensionless 
coefficient.

\section{Conclusion}

The purpose of this paper was to introduce the new approach to the cosmological
 model building based on $F(\car)$ supergravity, develop the techniques of
deriving $f(R)$-gravity from $F(\car)$ supergravity, and apply them to the 
simplest example (\ref{an}) with $F''(\car)=const.\neq 0$. Our choice 
(\ref{an}) of the $F$-function was not physically motivated but was dictated by
 technical simplicity only. Already the simplest example (\ref{an}) reveals 
several important new features that are superior to those of the usual 
supergravity characterized by $F''(\car)=0$, namely,
\begin{itemize} 
\item not any $f(R)$-gravity is extendable to $F(\car)$ supergravity; a simple
 (polynomial) choice of the $F(\car)$-function leads to a complicated 
(non-polynomial) bosonic $f(R)$-function;
\item it is easy to get a {\it positive} cosmological constant, see eg., 
eq.~(\ref{cc});  it may have physical applications to (primordial) dark energy;
\item the natural appearance of the (AdS) {\it bound} on the scalar curvature, 
resembling the Special Relativity bound on the physical speed, and similar to 
the Born-Infeld bound on the maximal values of the electro-magnetic field 
strength; 
\item the existence of instabilities;
\item removing the obstacles for a simple and natural (F-term-type) realization
 of the cosmological inflation in supergravity \cite{ks}.
\end{itemize} 
As regards the last statement, some comments are in order. As is well known 
\cite{llbook},
there is a generic problem of realizing inflation in supergravity, the 
so-called $\eta$-problem \cite{yana1}. To avoid the $\eta$-problem, one usually
 assumes that the K\"ahler potential of the chiral matter does not depend upon 
some chiral superfields (called flat directions), while the inflaton is 
supposed to be associated with one of the flat directions \cite{yana2,kl1}. 
Though it is possible to realize inflation that way, the mechanism is 
apparently non-geometrical and requires extra fields beyond supergravity. On 
the other side, the $f(R)$-gravity realization of inflation is easy and 
geometrical \cite{fr,stu}, there exist its very economical formulation 
\cite{star}, while it can also be extended to $F(\car)$ supergravity \cite{ks}.
  
Since $F(\car)$ supergravity is classically equivalent to some standard chiral 
matter-coupled supergravity \cite{our1}, one may wonder, how $F(R)$ 
supergravity avoids the $\eta$-problem? The basic explanation is that the 
K\"ahler potential, which follows from $F(\car)$ supergravity, appears to be 
non-trivial (ie. is given by a non-linear sigma-model),  whereas its naive
(free) form was one of the main assumptions used in the $\eta$-problem 
derivation  \cite{yana1}. 

Moreover, it is difficult to analyze the stability conditions and compute the 
spectral indices of inflation in any matter-coupled supergravity model with a 
non-trivial K\"ahler potential of matter.~\footnote{See, however, a recent 
limited analysis in ref.~\cite{stanf}.} Therefore, the possibility of a 
relatively simple analysis on the $F(\car)$ supergravity side is yet another 
advantage of our approach. 

Both the Legendre-Weyl transform in $f(R)$ gravity and the 
Legendre-Weyl-K\"ahler transform in $F(\car)$-supergravity, needed to prove 
their equivalence to the scalar-tensor gravity (quintessence) and the standard 
matter-coupled supergravity, respectively, apply at the classical level, while 
they are likely to fail at the quantum level, because of the 
non-renormalizability on the quantum field theories on both sides.  Revealing 
the (quantum) origin of the higher-order scalar supercurvature terms  in the 
supergravity function $F(\car)$ (say, from some fundamental theory of Quantum 
Gravity, like Superstrings/M-theory) is beyond the scope of this paper and 
beyond our ability at present.

We do not exclude possible applications of $F(\car)$ supergravity to the 
{\it present} Universe and the {\it present} Dark Energy, by choosing an 
appropriate function $F(\car)$. However, one needs more physical reasoning for 
applying supergravity to a low-curvature cosmology. In its turn, it requires 
knowing more details about supersymmetry breaking. The example (\ref{an}) 
considered in this paper does not have the correct Newton limit, unless 
the first and last terms in eq.~(\ref{an}) are negligible.

\section*{Acknowledgements}

One of the authors (SVK) is grateful to H. Abe, L. Amendola, G. Dvali, 
A. De Felice, A. Hebecker, S. Hellerman, T. Kobayashi, K.-I. Maeda, 
S. Mukohyama, M. Sasaki, A.A. Starobinsky, S. Tsujikawa and J. Yokoyama for 
discussions.

\end{document}